\documentclass[intlimits,twoside,a4paper]{article}

\usepackage[cp1251]{inputenc}

\usepackage[eqsecnum]{cmpj3}

\issue{2025}{28}{1}{13501}
\doinumber{10.5488/CMP.28.13501}

\title[Advances of cluster approach for lattice models]%
{Advances of Mayer's cluster approach in quantitative theoretical description of phase transitions for various lattice models of matter}
\author[M. V. Ushcats \textit{et al.}]{
M. V. Ushcats \orcid{0000-0002-0174-1594}\refaddr{addr1}\thanks{Corresponding author: \email{mykhailo.ushcats@nuos.edu.ua}},
        L. A. Bulavin\orcid{0000-0002-8063-6441}\refaddr{addr2}, 
        S. Yu. Ushcats
        \orcid{0000-0002-5250-4505}
        \refaddr{addr1}, 
        Zh. Yu. Burunina\orcid{0000-0001-7631-7213}\refaddr{addr1},
        O.~V.~Maiboroda\orcid{0000-0001-9206-4316}\refaddr{addr1},
        N. O. Romanchuk\orcid{0000-0002-3225-6428}\refaddr{addr1},
        N. O. Shapoval\orcid{0000-0002-5452-8148}\refaddr{addr1}
}
\addresses{
\addr{addr1} Admiral Makarov National University of Shipbuilding, 9 Heroes of Ukraine Ave., Mykolayiv 54025, Ukraine
\addr{addr2} Taras Shevchenko National University of Kyiv, 2 Academician Glushkov Ave., Kyiv 03680, Ukraine
} 

\date{Received November 19, 2024, in final form February 10, 2025}
\begin{document}

\maketitle

\begin{abstract}
Resent achievements in statistical theory, namely, a possibility to reproduce almost unlimited Mayer's activity series based on the information about their convergence radius, on the one hand, and generalization of the lattice statistics by eliminating the simplification of nearest-neighbor interactions, on the other hand, have allowed accurate quantitative description of the condensation in lattice gases, spontaneous magnetization in ferromagnets, and spinodal decomposition in binary mixtures by evaluating only  several irreducible cluster integrals (virial coefficients). In particular, the results of calculations indicate qualitative and even quantitative universality in the behavior of the mentioned lattice systems of different geometry and dimensionality at the same values of a certain reduced temperature when that behavior is expressed in terms of some dimensionless parameters. An additional possibility to describe the order-disorder phase transitions in some other lattice systems (e.g., antiferromagnets and alloys) is also discussed in the paper.
%
\end{abstract}

\section{Introduction}

One of the long-standing problems of molecular physics, physical chemistry, etc. is the theoretical statistically-based description of phase transitions in thermodynamic systems. For many years, there have been an extremely few achievements in this issue on the qualitative level for specific and very simplified models of matter only \cite{Kac2, Lebowitz2, LY}. Some recent generalizations based on the Mayer cluster expansion \cite{Mayer} are also qualitative rather than quantitative: namely, studies of the so-called virial series in terms of reducible and irreducible cluster integrals \cite{Mayer, JML} (virial coefficients specific for a certain interaction model) as well as new equations of state \cite{PRL, Bannur, PRE5} in terms of the same integrals have finally cleared the physical meaning of the regimes where the virial series diverge mathematically \cite{Lebowitz1, PRE4, PRE6, PRE1, UJP4}  and established a strict mathematical definition for the saturation point that is general for all classical fluids \cite{JML, PRE6, JCP1} and even quantum systems \cite{Suresh2018, Suresh2020}.

As to the quantitative description of phase transitions, the existing achievements remain much less significant: in practice, any accurate cluster-based definition of the saturation point for a certain statistical model of matter needs calculations of the corresponding cluster integrals to the orders of hundreds or thousands that can hardly be performed even by using the modern computational techniques and equipment. 

Nevertheless, an approximate technique was proposed recently \cite{PRE7}, which allows a pretty accurate evaluation of the saturation point even for various real substances by using the empirical information on only the second virial coefficient and phase-transition activity (fugacity): in fact, the latter quantity is directly related to the mathematical divergence of virial series and, therefore, it incapsulates the information on the unknown high-order cluster integrals (virial coefficients) \cite{Pramana}.

Concerning this technique, thermodynamic systems with the discrete configuration phase-space (the so-called lattice models of matter: lattice gases, the Ising model of magnetics \cite{Ising}, etc.) turn out to be of special interest in quantitative theoretical description of phase transitions based on the interaction parameters: on the one hand, the calculation of low-order cluster integrals is not a difficult problem (especially, for these models) and, on the other hand, the ``hole-particle'' symmetry of such models~\cite{Hill, PRE2, PRE3} allows an exact theoretical prediction of the phase-transition activity \cite{PRE6, UJP5, PHYSA1} without any additional empirical data. 

Another reason to renew the interest of statistical theory to lattice models is an essential recent generalization of such models due to eliminating the traditional simplification of nearest-neighbor (n.n.) interactions that has been an almost integral part of the lattice statistics (see chapter 7 in the famous book of T. Hill \cite{Hill}) for a long time. Today, the lattice model of any physical nature can actually include pairwise interactions on arbitrary distances between particles (in fact, the corresponding interaction potential may even be anisotropic \cite{PRE2, PHYSA1, PHYSA2}), thus potentially making such a model as realistic as possible.

In this short review, we present a detailed description of the above-mentioned latest achievements in the quantitative statistical theory of lattice systems, especially, concerning different phase transitions in such systems of various geometry and physical nature: phenomena of condensation in lattice gases, magnetization of ferromagnets, and spinodal decomposition in binary mixtures. First (section~\ref{sec2}), the theoretical background of the modern approximation technique is described as well as the new formal mathematical relationship among physically distinct lattice models. In section~\ref{sec3}, the results of calculations are presented for those distinct models of various geometry. Finally, those results and further perspectives of the approach are discussed in the last section.

\section{Theoretical background}\label{sec2}

\subsection { Mayer's expansion and evaluation of cluster integrals }

Mayer's expansions \cite{Mayer,JML} for pressure ($P$) and particle number density ($ \rho = N / V$) in powers of \emph{activity} (or, sometimes, \emph{fugacity}),

\[
z = {\lambda ^{ - 3}}\exp \left( {\frac{\mu }{{{k_\text{B}}T}}} \right) ,
\]
where $\mu $ is the chemical potential and $\lambda = h/\sqrt {2\pi m{k_\text{B}}T} $ is the thermal wavelength, have the following general form:

\begin{equation}
\left. \begin{array}{l}
\displaystyle \frac{P}{{{k_\text{B}}T}} = \sum\limits_{n = 1}^\infty  {{b_n}{z^n}} \\
\displaystyle \rho  = \sum\limits_{n = 1}^\infty  {n{b_n}{z^n}} 
\end{array} \right\} \label{eq:1},
\end{equation}
which is valid for continuous systems as well as lattice models of matter in all their subcritical gaseous regimes \cite{PRL, JML, PRE4, PRE5, PRE6}: from dilute (ideal) gases (when $z \to 0$) up to the condensation point, $ z = {z_{S}}$ [${z_{S}}$ --- the convergence radius for the power series in (\ref{eq:1})].

For lattice models exclusively, the ``hole-particle'' symmetry of their partition function yields the following ``high-density'' expansions for pressure and density \cite{PRE2, PRE3}:

\begin{equation}
\left. \begin{array}{l}
\displaystyle  \frac{P}{{{k_\text{B}}T}} = {\rho _0}\left( {\frac{{{u_0}}}{{{k_\text{B}}T}} + \ln \frac{{{\rho _0}}}{ \eta }} \right) + \sum\limits_{n \geqslant 1} {{b_n}{ \eta ^n}}  \\ 
\displaystyle  \rho  = {\rho _0} - \sum\limits_{n \geqslant 1} {n{b_n}{ \eta ^n}}  \\ 
 \end{array} \right\} \label{eq:2} ,
\end{equation}
in powers of the ``reciprocal activity'',

\begin{equation}
\eta = \frac{{\rho _0^2}}{z}\exp \left( {2\frac{{{u_0}}}{{{k_\text{B}}T}}} \right) \label{eq:3} ,
\end{equation}
where $u_0$ is the potential energy per particle in the \emph{close-packing state} and $ \rho_0$ is the particle number density in such close-packing state. It is obvious that the convergence radii are identical for the power series of equations~(\ref{eq:1}) and (\ref{eq:2}) (i.e., $ {\eta_{S}} \equiv {z_{S}}$, where $ \eta_{S} $ takes the meaning of the ``boiling reciprocal activity'').

Analytically, it is easy to check (no matter which certain set of power coefficients, $\left \{ b_{n} \left ( T \right ) \right \}$) that equations~(\ref{eq:1}) and (\ref{eq:2}) must always define the equal values of pressure at $ {\eta_{S}} = {z_{S}}$ [i.e., $ P ( {z_S} ) \equiv P ( {\eta_{S}} ) $] and symmetrical values of density [$ \rho ( {z_S} ) = { \rho_{0}} - \rho ( {\eta_{S}} ) $]  that, at subcritical temperatures [$ \rho ( {z_S} ) < { \rho_{0}} / 2 $], exactly corresponds to the first-order phase transition --- the jump of density at a constant pressure and activity (chemical potential) \cite{PRE6}.

In fact, the accuracy of equations~(\ref{eq:1}), (\ref{eq:2}) explicitly depends on the accuracy of the power coefficients, $\left \{ b_{n} \left ( T \right ) \right \}$, belonging to both equations. These coefficients are called \emph{reducible cluster integrals}~\cite{Mayer, JML} and their evaluation for some more or less realistic thermodynamic system (any actual $\left \{ b_{n} \right \}$ set must include the infinite number of reducible cluster integrals) is the main problem in practical usage of equations~(\ref{eq:1}), (\ref{eq:2}).

As a rule, the \emph{irreducible integrals}, $\left \{ \beta _{k} \left ( T \right ) \right \}$, (the corresponding \emph{virial coefficients}) are calculated for a given interaction model and then, based on the accurate and complete (i.e., infinite) $\left \{ \beta _{k} \right  \}$ set, one could easily obtain the accurate and complete (infinite) $ \{ b_{n} \}$ set by using the following recursive algorithm~\cite{PRE4}:

\begin{equation}
{b_n} = \frac{A_{{n},{n - 1}}}{n^2}, \label{eq:4}
\end{equation}
where
\[{A_{n,i}} = n\sum\limits_{k = 1}^i {\frac{k}{i}{\beta _k}{A_{{n},{i - k}}}}; \]  
\[{A_{n,0}} = 1.\]

In addition, the convergence radius for the power series of equation~(\ref{eq:1}) could be defined on the basis of the known $ \{ \beta _{k} \}$ set:
\begin{equation}
{z_S} = {\rho _S}\exp \left( { - \sum\limits_{k \geqslant 1} {{\beta _k}\rho _S^k} } \right) , \label{eq:5}
\end{equation}
where the saturation density, $\rho _S$, is the minimum positive root in
\begin{equation}
\sum\limits_{k \geqslant 1} {k{\beta _k}\rho _S^k}  = 1 . \label{eq:5a}
\end{equation}

At the moment, there is a great experience in calculations of virial coefficients up to the 7th--16th orders for various interaction models \cite{KofkePRL, UJP1, UJP2, JCP2, UJP3, Wheatley, Kofke5, CMP1}, but the neglect of the other coefficients (of higher orders) dramatically affects the accuracy of the corresponding equations at the vicinity of the phase-transition point. Though, at the first glance, even the infinite $ \{ b_{n} \}$ set can be reproduced by the recursive relation in equation~(\ref{eq:4}) on the basis of such extremely limited $ \{ \beta _{k} \}$ sets, the accuracy of the resulting  $ \{ b_{n} \}$ sets in high orders is unfortunately very poor and nonsatisfactory as well as the accuracy of the corresponding values of $z_S$ and $\rho _S$ in equations~(\ref{eq:5}), (\ref{eq:5a}). In fact, any more or less accurate cluster-based evaluation of the saturation point requires calculations of irreducible integrals to the orders of hundreds or thousands~\cite{JCP3, Kofke6} which is technically impossible today.

As to the lattice models, there is another option to define $z_S$ exactly \cite{PRE6} based on the above-mentioned parameters, $u_0$ and $ \rho_0$, of the lattice model [see equations~(\ref{eq:2}), (\ref{eq:3})]: 
\begin{equation}
z_S \equiv \eta _S = {\rho _0} \exp \left( {\frac{{u_0}}{{k_\text{B}}T}} \right) \label{eq:6} .
\end{equation}

In turn, that $z_S$ value defines the asymptotic behavior of high-order cluster integrals in the correct~$\{ b_{n} \}$ set \cite{Pramana} and, thus, allows their approximate evaluation. For example, the simplest approach to such approximation is a trivial scaling \cite{UJP5} of all the ``incorrect'' reducible integrals (if the initial set of the calculated irreducible integrals, $ \{ \beta _{k} \}$, is limited by the maximum order of $k_\text{max}$, the reducible integrals,~$ \{ b_{n} \}$, calculated by using the recursive relation in equation~(\ref{eq:4}) on the basis of that limited~$\{ \beta _{k} \}$ set can be considered as correct up to the order of ${n_\text{max}} = {k_\text{max}} + 1$ only, whereas all the other integrals of higher orders cannot be considered as correct). Namely, 
\begin{equation}
b^{\text{true}}_{n} = b_{n} {\left( {\frac{{z_S}}{z^{\text{true}}_{S}}} \right)}^{n-1} \label{eq:7} ,
\end{equation}
where $ b^{\text{true}}_{n}$ is a "corrected" reducible integral ($n > {k_\text{max}} + 1$), $b_{n}$ and $ z_S$ are ``incorrect'' values (calculated on the basis of the limited $ \{ \beta _{k} \}$ set), and $z^{\text{true}}_{S}$ is the true convergence radius defined by equation~(\ref{eq:6}).

In \cite{PRE7, PRE8}, a more complex approximation technique is considered, where the transition from the ``correct'' low-order integrals ($n \leqslant {k_\text{max}} + 1$) to the modified high-order integrals ($n \gg {k_\text{max}} + 1$) is additionally smoothed, although in case of lattice models, such smoothing has almost  no effect on the results \cite{PHYSA1}.

\subsection { Relationship among the parameters of lattice gases, ferromagnets, and binary mixtures }

In the paragraphs above, Mayer's expansions in equations~(\ref{eq:1}), (\ref{eq:2}) concern rather the lattice models of classical fluids (the so-called \emph{lattice gases} with the discrete configurational phase-space of $N_0$ cells, where one space cell cannot be occupied by more than a single particle) and are applicable at subcritical regimes to describe condensation/vaporization phenomena only.

On the other hand, there exists a well-known formal relationship of such lattice-gas models with other specific statistical models of matter: the Ising model of ferromagnets \cite{Ising} and simplified models of binary mixtures/solutions/alloys. In details, the derivation of this relationship can be found in a number of sources \cite{Hill, LY, PHYSA1} and it is based on the mathematical similarity of the partition functions for all the mentioned models with the discrete configuration phase-space.

Unfortunately, all the mentioned lattice models of matter have traditionally involved a principal restriction: any particle of a lattice gas (a spin of a magnet or a molecule of a mixture) can only interact by some constant strength with its \emph{nearest neighbors} in the closest cells of the space lattice (the so-called n.n. or nearest-neighbor lattice statistics \cite{Hill}), and such restriction greatly affected the actual applicability of lattice models in describing realistic interactions.

At first, this n.n. restriction has been eliminated recently for lattice-gas models \cite{ PRE2, PRE3}. Namely, the Hill approach to lattice statistics \cite{Hill} was reformulated in terms of the general pairwise interaction potential, $ \phi \left ( {\mathbf{r}_i}, {\mathbf{r}_j} \right ) $, depending on positions, ${\mathbf{r}_i}$ and ${\mathbf{r}_j}$, of any pair of molecules, $i$ and $j$, anywhere in the lattice (the $ \phi$ function is positive for repulsion and negative for attraction and, in fact, it is not required to be spherically symmetrical). Based on the $ \phi$ function, the $u_0$ parameter was introduced [see equation~(\ref{eq:2})] and the radius of convergence of Mayer's activity expansions (i.e., the phase-transition activity), $z_S$ , was established \cite{PRE6} [see equation~(\ref{eq:6})].

\begin{table}[t]
\noindent\caption{\label{tab:1} Relationship among the parameters of lattice-gases, the Ising model of magnetics, and the lattice model of binary mixtures}
\vskip3mm\tabcolsep4.5pt

\begin{center}
\renewcommand{\arraystretch}{2.0}

\begin{tabular}{||c|c|c||}

\hline 
\hline 
   \textbf{lattice gas} & \textbf{Ising model} & \textbf{binary mixture} \\[6pt]
\hline
\hline 
   $\phi \left ( {\mathbf{r}_i}, {\mathbf{r}_j} \right )$ & $4 \Phi \left ( {\mathbf{r}_i}, {\mathbf{r}_j} \right )$ & $ \phi '({{\bf{r}}_i},{{\bf{r}}_j}) = {\phi _{aa}} + {\phi _{bb}} - 2{\phi _{ab}} $  \\[6pt] 
\hline 

   $ z = {\lambda ^{ - 3}}\exp \left( {\frac{\mu }{{{k_\text{B}}T}}} \right)$ & $ {\rho _0} \exp \left ( \frac {2H + {u_0}} {{k_\text{B}} T}  \right )$ & $ z' = {\rho _0}\exp \left( {\frac{{{u_0}}}{{{k_\text{B}}T}}} \right)\frac{{{z_a}\exp \left( { - \frac{{{u_a}}}{{{k_\text{B}}T}}} \right)}}{{{z_b}\exp \left( { - \frac{{{u_b}}}{{{k_\text{B}}T}}} \right)}}$  \\[6pt]
\hline 

   $\frac {P} {\rho _0}$ & $ H + \frac {u_0}{4} - F - {k_\text{B}} T \ln { \frac {j_m}{\rho _0} }$ & $- {{k_\text{B}}T \ln {Z_b}} = {u_b} - { \mu _b} - {{k_\text{B}}T \ln {j_b}}$ \\[6pt]
\hline 

   $\frac {\rho} {\rho _0}$ & $\frac {1+I} {2}$  & ${n_a} = \frac{{{N_a}}}{{{N_a} + {N_b}}}$   \\[6pt]

\hline 
\hline 
\end{tabular}

\renewcommand{\arraystretch}{1.0}

\end{center}
\end{table}

Later, the same formalism was confirmed for the Ising models \cite{PHYSA1}. For a ferromagnet in an external magnetic field of some intensity, $H$, all the $N_0$ space cells are actually occupied by spins (one spin per cell) and each spin can have only one of the two orientations: either in the direction of the external field (parallel or formally positive spins, the number of which can be denoted as $N^{+}$) or in the opposite direction (antiparallel or negative spins, the number of which is $N^{-} = {N_0} - N^{+}$). In the partition function, the integration over the non-configurational phase-space yields some quantity $ j^{N_0}_m $ (i.e., $ {j_m} $ per spin), whereas the configuration integral (actually, some sum for the discrete configurational phase-space) may be represented in a form mathematically analogous to that of lattice-gas models \cite{ PHYSA1}. Again, depending on the orientation of any two spins, $i$ and $j$, at some arbitrary positions in the space lattice, ${\mathbf{r}_i}$ and ${\mathbf{r}_j}$, their interaction energy may be described by a certain $ \Phi \left ( {\mathbf{r}_i}, {\mathbf{r}_j} \right ) $ function formally related to the $ \phi \left ( {\mathbf{r}_i}, {\mathbf{r}_j} \right ) $ of lattice-gas models (see table~\ref{tab:1}). 

As a result, the magnetic free energy per spin, $F$, and dimensionless intensity of magnetization, $I = (N^{+} - N^{-})/(N^{+} + N^{-})$, in the Ising problem of some geometry and dimensionality can be formally related to the pressure and density of the corresponding lattice-gas model (see table~\ref{tab:1}). 

In a mixture model, each cell is occupied by a molecule belonging to one of the two components: $a$ or $b$. There, the situation seemed to be somewhat problematic because the presence of three different and independent interaction potentials (namely, the $\phi _{aa} \left ( {\mathbf{r}_i}, {\mathbf{r}_j} \right )$ for $a$-$a$ interactions, $\phi _{bb} \left ( {\mathbf{r}_i}, {\mathbf{r}_j} \right )$ for $b$-$b$ interactions, and $\phi _{ab} \left ( {\mathbf{r}_i}, {\mathbf{r}_j} \right )$ for $a$-$b$ interactions) essentially complicates the partition function in comparison with the conventional n.n. models (which include only the corresponding constant energy values of nearest-neighbor interactions).

Nevertheless, the last efforts \cite{PHYSA2} to express the mixture partition function in terms of $\phi _{aa}$, $\phi _{bb}$, $\phi _{ab}$ in a form analogous to that of the Ising and lattice-gas models have led to a success and the above-mentioned generalized relationship between the Ising and lattice-gas models was finally extended to the case of mixtures.

In such formalism, the \emph{reduced interaction potential},
\begin{equation}
\phi '({{\bf{r}}_i},{{\bf{r}}_j}) = {{\phi _{aa}}({{\bf{r}}_i},{{\bf{r}}_j}) + {\phi _{bb}}({{\bf{r}}_i},{{\bf{r}}_j})} - 2{\phi _{ab}}({{\bf{r}}_i},{{\bf{r}}_j}), \label{eq:8}
\end{equation}
is considered as an analogue of the lattice-gas $ \phi \left ( {\mathbf{r}_i}, {\mathbf{r}_j} \right ) $ function (namely, all the cluster integrals in equations~(\ref{eq:1}), (\ref{eq:2}), (\ref{eq:4}), (\ref{eq:5}), (\ref{eq:7}) as well as the $u_0$ parameter are defined by this reduced potential), whereas the $\phi _{aa}$ and $\phi _{bb}$ functions remain incapsulated in the corresponding ${u_a}$ and ${u_b}$ parameters: ${u_a}$ is the potential energy per molecule $a$ when all its neighbors (in all coordination spheres) are also $a$ molecules; ${u_b}$ is the similar value at the close-packing state of $b$ molecules only.

Furthermore, those parameters, ${u_a}$, ${u_b}$, and per-component activities,
\[
{z_a} = {{\lambda_{a}^{ - 3}}}\exp \left( {\frac{\mu _a }{{{k_\text{B}}T}}} \right),
\]
\[
{z_b} = {{\lambda_{b}^{ - 3}}}\exp \left( {\frac{\mu _b }{{{k_\text{B}}T}}} \right),
\]
allow introducing the mixture analogue of the lattice-gas activity
\begin{equation}
z' = {\rho _0} {\frac{Z_a}{Z_b}} \exp \left( {\frac{{{u_0}}}{{{k_\text{B}}T}}} \right), \label{eq:9}
\end{equation}
where 
\begin{equation}
\left. \begin{array}{l}
{Z_{a}} = {z_{a}}\exp \left( { - \dfrac{{{u_{a}}}}{{{k_\text{B}}T}}} \right)\\
{Z_{b}} = {z_{b}}\exp \left( { - \dfrac{{{u_{b}}}}{{{k_\text{B}}T}}} \right) 
\end{array} \right\} \label{eq:10}.
\end{equation}

As a result, the following quantity of a binary mixture:
\[
{u_b} - { \mu _b} - {{k_\text{B}}T \ln {j_b}} = - {{k_\text{B}}T \ln {Z_b}},
\]
becomes a formal analogue of pressure in the corresponding lattice-gas model, whereas the relative concentration (molar fraction) of the $a$ component,
\[
{n_a} = \frac{{{N_a}}}{{{N_a} + {N_b}}},
\]
is a dimensionless analogue of the lattice-gas density (see table~\ref{tab:1}).

\section{Numerical application}\label{sec3}

\subsection { Condensation of lattice gases }

As it is mentioned above, the ``low-density'' expansions in equation~(\ref{eq:1}) and ``high-density'' expansions in equation~(\ref{eq:2}) can be used directly in order to construct a theoretical isotherm, $P \left( \rho \right)$, of some lattice-gas model at a given temperature, $T$, under the following condition: the $ \{ {b_n} \} $ set of reducible cluster integrals is known to very high orders (to thousands) with more or less accuracy for the model at this temperature. 

\begin{table}[htb]
\noindent\caption{\label{tab:2}Irreducible cluster integrals for the two-dimensional square model, where each particle can interact with four nearest neighbors (the Lee-Yang model, $ {u_0} = - 2 \varepsilon $).
}\vskip3mm\tabcolsep4.5pt

\centering{

\renewcommand{\arraystretch}{1.3}
\noindent{\footnotesize\begin{tabular}{||c|l||}
\hline 
\hline
\rule{0pt}{3mm}
$k$ & $k \beta _k \rho _{0} ^{k} $

\\
\hline 
\hline

$1$ & $4f - 1$\\[1mm]
$2$ & $- 12{f^2} - 1$\\[1mm]
$3$ & $12{f^4} + 40{f^3} + 12{f^2} - 1$\\[1mm]
$4$ & $- 160{f^5} - 220{f^4} - 80{f^3} - 1$\\[1mm]
$5$ & $60{f^7} + 1380{f^6} + 1704{f^5} + 600{f^4} + 40{f^3} - 1$\\[1mm]
$6$ & $- 1428{f^8} - 10584{f^7} - 13440{f^6} - 5376{f^5} - 588{f^4} - 1$\\

\hline
\hline 
\end{tabular}}
\renewcommand{\arraystretch}{1.0}

}

\end{table}

\begin{table}[htb]
\noindent\caption{\label{tab:3}Irreducible cluster integrals for the two-dimensional square model, where each particle can interact with eight nearest neighbors ($ {u_0} = - 4 \varepsilon $).
}\vskip3mm\tabcolsep4.5pt

\centering{

\renewcommand{\arraystretch}{1.3}
\noindent{\footnotesize\begin{tabular}{||c|l||}
\hline 
\hline
\rule{0pt}{3mm}
$k$ & $k \beta _k \rho _{0} ^{k} $

\\
\hline 
\hline

$1$ & $8f - 1$\\[1mm]
$2$ & $24{f^3} - 24{f^2} - 1$\\[1mm]
$3$ & $12{f^6} + 168{f^5} - 144{f^4} - 64{f^3} + 24{f^2} - 1$\\[1mm]
$4$ & $180{f^8} + 1120{f^7} - 1760{f^6} - 1280{f^5} + 680{f^4} + 80{f^3} - 1$\\[1mm]
$5$ & $180{f^{11}} + 2820{f^{10}} + 8460{f^9} - 17580{f^8} - 8940{f^7} + 17520{f^6} + 4728{f^5} - 1320{f^4} - 40{f^3} - 1$\\[1mm]
$6$ & $252{f^{14}} + 4872{f^{13}} + 32844{f^{12}} + 41664{f^{11}} - 216636{f^{10}} - 87192{f^9} + 218736{f^8} + 17976{f^7}$\\
$ $ & $ - 78960{f^6} - 9744{f^5} - 96940{f^4} - 81396{f^3} - 166488{f^2} - 15624{f} - 1$\\

\hline
\hline 
\end{tabular}}
\renewcommand{\arraystretch}{1.0}

}

\end{table}

\begin{table}[htb]
\noindent\caption{\label{tab:4}Irreducible cluster integrals for the two-dimensional triangle model, where each particle can interact with six nearest neighbors ($ {u_0} = - 3 \varepsilon $).}
\vskip3mm\tabcolsep4.5pt

\centering{

\renewcommand{\arraystretch}{1.3}
\noindent{\footnotesize\begin{tabular}{||c|l||}
\hline 
\hline
\rule{0pt}{3mm}
$k$ & $k \beta _k \rho _{0} ^{k} $

\\
\hline 
\hline 

$1$ & $6f - 1$\\
\hline 
$2$ & $12{f^3} - 18{f^2} - 1$
\\
\hline 
$3$ & $36{f^5} - 108{f^4} - 12{f^3} + 18{f^2} - 1 $
\\
\hline 
$4$ & $120{f^7} - 540{f^6} + 240{f^5} + 510{f^4} - 1$
\\
\hline 
$5$ & $420{f^9} - 2520{f^8} + 2790{f^7} + 3180{f^6} - 2304{f^5} - 990{f^4} - 1$\\

\hline
\hline 
\end{tabular}}
\renewcommand{\arraystretch}{1.0}

}

\end{table}

\begin{table}[htb]
\noindent\caption{\label{tab:5}Irreducible cluster integrals for the three-dimensional cubic model, where each particle can interact with six nearest neighbors ($ {u_0} = - 3 \varepsilon $).}
\vskip3mm\tabcolsep4.5pt

\centering{

\renewcommand{\arraystretch}{1.3}
\noindent{\footnotesize\begin{tabular}{||c|l @{\hspace{110pt}}||}
\hline 
\hline
\rule{0pt}{3mm}
$k$ & $k \beta _k \rho _{0} ^{k} $

\\
\hline 
\hline 

$1$ & $6f - 1$\\
\hline 
$2$ & $- 18{f^2} - 1$\\
\hline 
$3$ & $36{f^4} + 60{f^3} +18{f^2}- 1$\\
\hline 
$4$ & $- 480{f^5} - 450{f^4} - 120{f^3} - 1$\\

\hline
\hline 
\end{tabular}}
\renewcommand{\arraystretch}{1.0}

}

\end{table}

\begin{table}[htb]
\noindent\caption{\label{tab:6}Irreducible cluster integrals for the three-dimensional cubic model, where each particle can interact with twenty six nearest neighbors ($ {u_0} = - 13 \varepsilon $).}
\vskip3mm\tabcolsep4.5pt

\centering{

\renewcommand{\arraystretch}{1.3}
\noindent{\footnotesize\begin{tabular}{||c|l @{\hspace{70pt}}||}
\hline 
\hline
\rule{0pt}{3mm}
$k$ & $k \beta _k \rho _{0} ^{k} $

\\
\hline 
\hline 

$1$ & $226 f - 1$\\
\hline 
$2$ & $264 {f^3} - 78 {f^2} - 1$\\
\hline 
$3$ & $804 {f^6} + 8280 {f^5} + 3312 {f^4} + 1324 {f^3} + 78 {f^2} - 1$\\
\hline 
$4$ & $1120 {f^{10}} + 15040 {f^9} + 109980 {f^8} + 399280 {f^7} + 277360 {f^6} - 36000 {f^5} - 22990 {f^4} + 2120 {f^3} - 1$\\

\hline
\hline 
\end{tabular}}
\renewcommand{\arraystretch}{1.0}

}
\end{table}

In turn, the calculation of such $ \{ b_{n} (T) \} $ set is proposed to be realized in several steps. First, a limited set of the corresponding irreducible cluster integrals, $ \{ \beta _{k} (T) \} $,  is defined up to some order $k_\text{max}$. Namely, the irreducible integrals up to the 4th--6th orders are presented by tables~\ref{tab:2}, \ref{tab:3}, \ref{tab:4}, \ref{tab:5}, \ref{tab:6} in a form of the exact function, $ \beta _{k} (T) $, for a number of different 2D and 3D n. n. lattice models. Actually, each irreducible integral is expressed there as a polynomial of one Mayer's function,
\[
f \left( T \right) = \exp \left( { \frac {\varepsilon} {{k_\text{B}} T} } \right) - 1,
\]
where $\varepsilon$ is the depth of the "attraction well" at the first coordination sphere radius, $r_1$ (i.e., where the nearest-neighbor interaction potential $ \phi ({r_1}) = -\varepsilon$). 

\begin{figure}[!t]
	\centering{
		\includegraphics[width=0.48\textwidth]{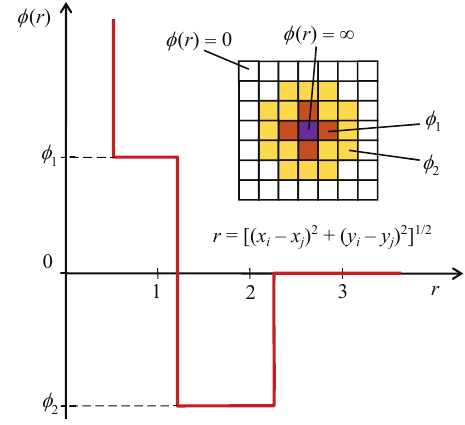}
	}
	\caption{(Colour online) ``Long-range interaction'' model --- the hard core interaction potential with different interactions in the first two coordination spheres.}
	\label{fig:1_1}
\end{figure}
In contrast to those n.n. models, table~\ref{tab:7}  presents the irreducible integrals for the 2D square ``long-range interaction'' model with interactions \emph{in two closest coordination spheres} (see figure~\ref{fig:1_1}) as polynomials of two Mayer's functions, $f_1$ and $f_2$ corresponding to the values of interaction energy, $ \phi _1 $ and $ \phi _2$, respectively.

\begin{table}[htb]
\noindent\caption{\label{tab:7}Irreducible cluster integrals for the two-dimensional ``long-range interaction'' model ($ {u_0} =  2 {\phi _1} + 8 {\phi _2}$, see figure~\ref{fig:1_1}).}
\vskip3mm\tabcolsep4.5pt

\centering{

\renewcommand{\arraystretch}{1.3}
\noindent{\footnotesize\begin{tabular}{||c|l||}
\hline 
\hline
\rule{0pt}{3mm}
$k$ & $k \beta _k \rho _{0} ^{k} $

\\
\hline 
\hline 

$1$ & $4{f_1}  + 16{f_2} - 1$\\
\hline 
$2$ & $ (36{f_2} - 12 ){f_1^2} + 96{f_2^2}{f_1} + 72{f_2^3}- 48{f_2^2} - 1$\\
\hline 
$3$ & $ (12{f_2^2}  + 24{f_2} + 12){f_1^4} + (192{f_2^3} + 648{f_2^2} - 144{f_2} + 40){f_1^3} + (216{f_2^4}+ 2112{f_2^3} + 240{f_2^2} - 216{f_2} + 12){f_1^2}$\\
$ $ & $+ (240{f_2^5}+ 2400{f_2^4}+ 1104{f_2^3} + 576{f_2^2}){f_1} + 12{f_2^6} + 792{f_2^5} + 336{f_2^4} - 272{f_2^3} + 448{f_2^2} - 1$\\
\hline 
$4$ & $(160{f_2^5} + 800{f_2^4} + 1600{f_2^3} + 1280{f_2^2} + 160{f_2} - 160){f_1^5}$ \\ 
$ $ & $ + (340{f_2^6} + 3480{f_2^5} + 12620{f_2^4} + 16280{f_2^3} - 1620{f_2^2} + 320{f_2} - 220){f_1^4}$\\
$ $ & $ + (320{f_2^7} + 4960{f_2^6} + 30080{f_2^5} + 67200{f_2^4} + 10800{f_2^3} - 72000{f_2^2} + 960{f_2} - 80){f_1^3}$\\
$ $ & $ + (160{f_2^8} + 3320{f_2^7} + 28560{f_2^6} + 106520{f_2^5}  + 55680{f_2^4} - 18720{f_2^3} - 1600{f_2^2} + 360{f_2}){f_1^2}$\\
$ $ & $ + (880{f_2^8} + 13280{f_2^7} + 65840{f_2^6} + 158400{f_2^5} - 15840{f_2^4} - 7360{f_2^3} + 960{f_2^2}){f_1}$\\
$ $ & $ + 980{f_2^8} + 14680{f_2^7} + 19160{f_2^6} - 1440{f_2^5} - 2800{f_2^4} + 400{f_2^3} - 1$\\

\hline
\hline 
\end{tabular}}
\renewcommand{\arraystretch}{1.0}

}

\end{table}

After defining such a limited $ \{ \beta _{k} (T) \} $ set, the first approximation for the $ \{ {b_n} \} $ set is calculated by using the recursive relation of equation~(\ref{eq:4}) to the orders as high as possible and, additionally, the approximate value of $z_S$ is defined by equation~(\ref{eq:5}) on the basis of the same initial $ \{ \beta _{k} \} $ set. 

Then, the ``incorrect'' reducible integrals ($ n > k_\text{max} + 1$) of that very approximate $ \{ {b_n} \} $ set are scaled by using equation~(\ref{eq:7}) on the basis of the calculated approximate $z_S$ and ``true'' convergence radius, $ z^{\text{true}}_{S}$, exactly defined by the model parameters in equation~(\ref{eq:6}). Finally, the resulting $ \{ b^{\text{true}}_{n} \} $ set can directly be used in calculations of pressure and density in equation~(\ref{eq:1}) (for gaseous regimes) and equation~(\ref{eq:2}) (for condensed regimes).

\begin{figure}[!t]
\centerline{\includegraphics[width=0.5\textwidth]{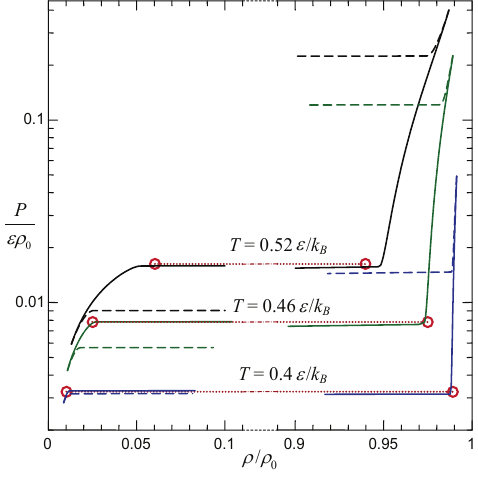}}
\caption{(Colour online) Isotherms of equation~(\ref{eq:1}) (left-hand curves) and equation~(\ref{eq:2}) (right-hand curves) with the unmodified $ \{ {b_n} \} $ (dashed curves) and scaled $\{ b^{\text{true}}_{n} \} $ (solid curves) sets of 1000 reducible cluster integrals defined on the basis of the $ \{ \beta _{1}, \beta _{2}, \beta _{3}, \beta _{4}, \beta _{5} \} $ sets at various temperature values of the two-dimensional square lattice gas (the Lee-Yang model). Circles correspond to the exact condensation parameters \cite{LY}.}
\label{fig:1_2}
\end{figure}

Figure~\ref{fig:1_2} clearly demonstrates the difference in accuracy of equations~(\ref{eq:1}), (\ref{eq:2}) with the mentioned $ \{ {b_n} \} $ and $ \{ b^{\text{true}}_{n} \} $ sets (both sets contain $1000$ integrals). There, the dashed lines show the isotherms of the Lee-Yang model (two-dimensional square lattice gas \cite{LY}) obtained by using equation~(\ref{eq:1}) (low-density regimes) and equation~(\ref{eq:2})  (high-density regimes) with the unmodified $ \{ {b_n} \} $ sets calculated on the basis of the $ \{ \beta _{1}, \beta _{2}, \beta _{3}, \beta _{4}, \beta _{5} \} $ sets at three different subcritical values of temperature (the critical temperature $T_C \approx 0.5673 \varepsilon / {k_\text{B}}$ for this model). In comparison with those dashed lines, the solid lines of figure~\ref{fig:1_2} (the isotherms of the same equations with the scaled $ \{ b^{\text{true}}_{n} \} $ sets) are much closer to the exact solution \cite{LY} (the circles in the figure) and, in particular, indicate almost identical pressure at the divergence regions of equations~(\ref{eq:1}), (\ref{eq:2}) --- regions corresponding to the gas-liquid phase transitions at the same three values of temperature.

\begin{figure}[!t]
\centerline{\includegraphics[width=0.5\textwidth]{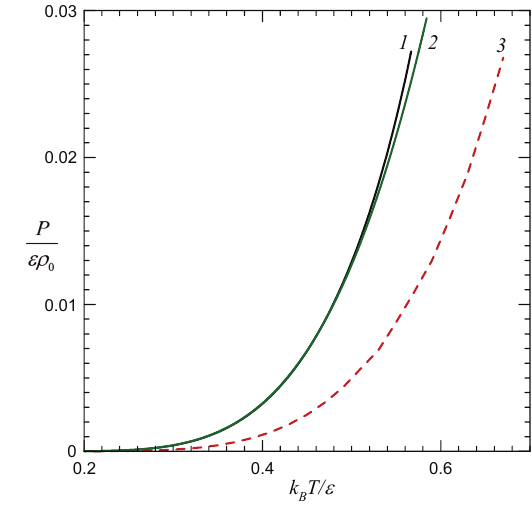}}
\caption{(Colour online) $P( T )$ binodal of the two-dimensional square lattice gas (the Lee-Yang model): 1 --- the exact solution \cite{LY}; 2 --- cluster expansions with $ \{ b^{\text{true}}_{n} \}$ sets of 1000 integrals defined on the basis of the $ \{ \beta _{1}, \beta _{2}, \beta _{3}, \beta _{4}, \beta _{5} \} $ sets at various temperature values; 3 --- the Curie-Weiss mean-field approximation \cite{Weiss}.}
\label{fig:1_4}
\end{figure}

In addition, a good agreement with exact condensation parameters \cite{LY} at various subcritical temperatures is demonstrated in figure~\ref{fig:1_4} (see the solid lines close to each other). For comparison, the binodal of the conventional Curie-Weiss mean-field approximation \cite{Weiss} (dashed line) is also presented in the figure. The obvious consistency between the cluster expansion and exact solution testifies to the qualitative as well as quantitative appropriateness of the described approach in subcritical regimes.

It should be noticed that, for other lattice-gas models, there exists no exact solution and, hence, this approach can serve as a powerful analytical alternative to computer simulations.

\begin{figure}[!t]
\centerline{\includegraphics[width=0.5\textwidth]{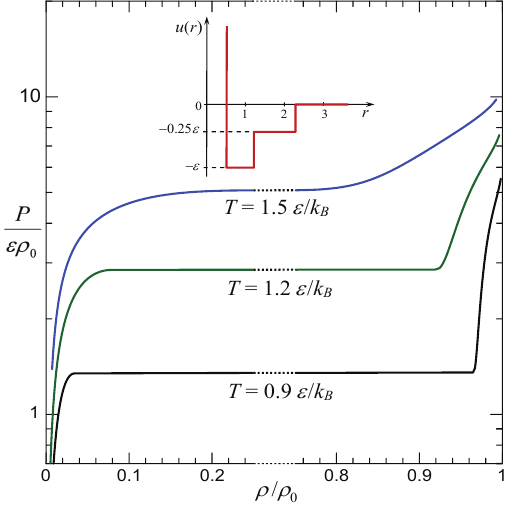}}
\caption{(Colour online) Isotherms of equation~(\ref{eq:1}) (left-hand curves) and equation~(\ref{eq:2}) (right-hand curves) for the 2D square ``long-range attraction'' lattice-gas model including  two different attraction wells in the first coordination sphere (the well depth is $\varepsilon$) and second coordination sphere (the well depth is $ \varepsilon / 4$), respectively.}
\label{fig:1_3}
\end{figure}

For example, figure~\ref{fig:1_3} shows the isotherms of a two-dimensional square lattice gas with the ``long-range interaction'' potential including two attraction wells of different depth (${ \phi _1} = - \varepsilon$, ${ \phi _2} = - \varepsilon / 4$), that we hereafter call the \emph{``long-range attraction'' potential}. In fact, all the differences among the subcritical isotherms of various lattice-gas models are quantitative rather than qualitative. The reader may compare the isotherms of figure~\ref{fig:1_3} with those presented in~\cite{UJP5} for another ``long-range interaction'' model, where four nearest particles are repulsed by each ``central'' particle (the barrier of height ${ \phi _1} = \varepsilon$ in the first coordination sphere) and the other sixteen particles are attracted (the attraction well of double depth in the second coordination sphere, ${ \phi _2} = - 2 \varepsilon$): the isotherms are quite similar qualitatively.

\subsection { Magnetization of ferromagnets }

In accordance with table~\ref{tab:1}, the density expansion of equation~(\ref{eq:1}) can be transformed to the following expression for the dimensionless intensity of magnetization of the Ising model with arbitrary geometry and dimensionality:
\begin{equation}
I = -1 + \frac{2}{\rho _0} { \sum\limits_{n = 1}^\infty  {n{b_n}{x^n}} } \label{eq:3_1} ,
\end{equation}
where the activity, $z$, is replaced by the $x$ variable,
\[
x = {\rho _0} \exp \left ( \frac {2H+{u_0}} {{k_\text{B}}T} \right ) ,
\]
and the $ \rho _0 $ is now defined by the volume, $ v_0 $, of a single cell of the ferromagnet: $ \rho _0 =  1 / v_0 $.

The physical and mathematical meaning of the reducible integrals, $\left \{ b_{n} \left ( T \right ) \right \}$, remains for the Ising model the same as for the lattice-gas of identical geometry and dimensionality but with the ``scaled'' interaction potential, $ \phi ( r )= 4 \Phi ( r )$  (see table~\ref{tab:1}).

It is important to note that, at subcritical temperatures (i.e., temperatures below the Curie point), the power series of equation~(\ref{eq:3_1}) converges only at $ x < {x_S}$, where $ x_S $ is equal to $ z_S $ defined by equation~(\ref{eq:6}). Moreover, the $x$ variable can be rewritten by using equation~(\ref{eq:6}):
\begin{equation}
x = {z_S} \exp \left ( \frac {2H} {{k_\text{B}}T} \right ) \label{eq:3_2}.
\end{equation}

It is easy to see that equation~(\ref{eq:3_1}) is valid only for negative magnetization ($ I < 0 $) at the correspondingly negative external field ($ H < 0 $ and, hence, $ x < {x_S} = {z_S} $). 
At a very strong negative external field ($ H \to -\infty$), magnetization is totally negative ($I \to -1$ at $x \to 0$) and its absolute value reduces when the absolute value of $H$ reduces.
Exactly at $ H = 0 $ ($x = {z_S}$), the power series of equation~(\ref{eq:3_1}) diverges from some point, $I \left( -0 \right) = -I_{m}$, in positive direction of $I$ and the value, $ I_{m}$, has the meaning of \emph{spontaneous magnetization}.

In order to obtain the positive branch of the magnetization curve, $ I \left ( H > 0 \right )$, the expansion for density in equation~(\ref{eq:2}) can be used:

\begin{equation}
I \left( H>0 \right) = 1 - \frac{2}{\rho _0} { \sum\limits_{n = 1}^\infty  {n{b_n}{y^n}} } \label{eq:3_3} ,
\end{equation}
where the $y$ variable is reciprocal to the above-mentioned $x$ and corresponds to the ``reciprocal activity'' of lattice gases in equation~(\ref{eq:3}), i.e.,
\[
y = \frac{{\rho _0^2}}{x}\exp \left( {2\frac{{{u_0}}}{{{k_\text{B}}T}}} \right),
\]
or, in accordance with equation~(\ref{eq:6}),
\begin{equation}
y = {z_S} \exp \left( \frac {-2H} {{k_\text{B}}T} \right) 
\label{eq:3_4}.
\end{equation}

Thus, a very strong positive external field ($ H \to \infty$) corresponds to the totally positive magnetization ($I \to 1$ at $y \to 0$) and when the field reduces, the magnetization reduces too approaching some specific value, $I \left( +0 \right) = I_{m}$, at $ H \to 0$, where the expression for $I$ in equation~(\ref{eq:3_3}) diverges in the negative direction that corresponds to the jump (phase transition) up to the above-mentioned negative branch of the magnetization curve. 

As a result, equation~(\ref{eq:3_1}) for $ I \left( H<0 \right) $ and equation~(\ref{eq:3_3}) for $ I \left( H>0 \right) $ define the intensity of magnetization as an odd function of $H$:
\[
I \left( -H \right) = -I \left( H \right).
\]

Such absolute symmetry of equations~(\ref{eq:3_1}) and (\ref{eq:3_3}) allows the practical usage of only one of them as a \emph{general solution of the Ising problem based on Mayer's cluster expansion}. Moreover, their formally distinct variables, $x$ and $y$, in equations~(\ref{eq:3_2}) and (\ref{eq:3_4}), respectively, can actually be considered as identical:
\[
x \equiv y = {z_S} \exp \left( \frac {-2 \left| H \right| } {{k_\text{B}}T} \right) .
\]

\begin{figure}
\centerline{\includegraphics[width=0.5\textwidth]{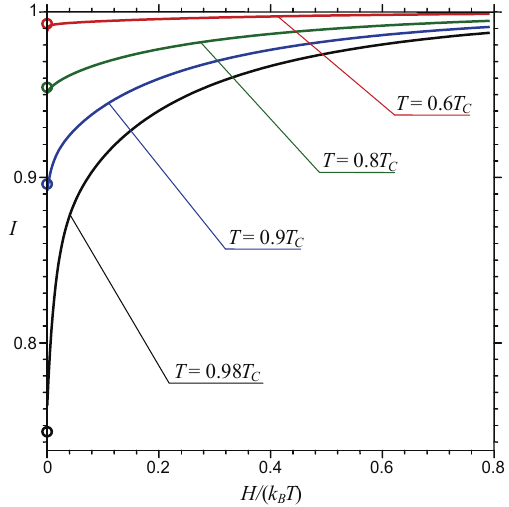}}
\caption{(Colour online) Magnetization curves of equation~(\ref{eq:3_3}) (i.e., positive branches) for the two-dimensional square Ising model (the Lee-Yang model). For each temperature, equation~(\ref{eq:3_3}) included the scaled $ \{ b^{\text{true}}_{n} \} $ set of 1000 reducible cluster integrals defined on the basis of the $ \{ \beta _{1}, \beta _{2}, \beta _{3}, \beta _{4}, \beta _{5} \} $ set. The circles correspond to the exact solution for the spontaneous magnetization \cite{Onsager, Kaufman, Yang}.}
\label{fig:2_1}
\end{figure}

For example, figure~\ref{fig:2_1} demonstrates the magnetization curves calculated for the two-dimensional square Ising model with nearest-neighbor interactions at four different subcritical temperatures. When $ H \to 0 $, those curves come close to the corresponding spontaneous magnetization points of the exact solution \cite{Onsager, Kaufman, Yang} known for this Ising model that, in turn, indicates the appropriateness of Mayer's expansion, in general, and the presented approximate approach to evaluate the large $\left \{ {b_n} \right \}$ set, in particular. Namely, changing the number of integrals in the $\left \{ b^{\text{true}}_{n} \right \}$ set (from $100$ to $10 000$) influences the results almost insignificantly. As to the number of irreducible integrals, $\left \{ {\beta _k} \right \}$, used to construct the $\left \{ b^{\text{true}}_{n} \right \}$ set of arbitrary length, its influence on the results is more significant: in order to calculate the reducible integrals for the curves of figure~\ref{fig:2_1}, the first five irreducible integrals are used whereas the decrease of this number leads to more considerable deviations of curves from the mentioned exact solution.

Unfortunately, the exact parameters of spontaneous magnetization are only known for the two-dimensional square Ising model with nearest-neighbor interactions that cannot help in testing the accuracy of approximated Mayer's expansion for other lattice models. Nevertheless, the usage of equation~(\ref{eq:3_3}) for a number of 2D and 3D models yields the results quite similar (at the qualitative level at least) to those presented in figure~\ref{fig:2_1} and, in detail, this similarity is considered in further sections.

\subsection { Spinodal decomposition in binary mixtures } 

In accordance with table~\ref{tab:1}, the lattice-gas expansions of equation~(\ref{eq:1}) for pressure, $P$, and density, $\rho$, in powers of activity, $z$, turn into similar expansions of $ \ln {Z_b} $ [see equation~(\ref{eq:10})] and $n_a$ (molar fraction) in powers of $z'$ [see equation~(\ref{eq:9})] for a lattice model of some binary mixture. In $z'$, the model parameter, $u_0$, must now be defined as the potential energy per one mixture particle as if all the particles interact with the reduced $ \phi ' ( r ) $ potential [see equation~(\ref{eq:8})] no matter to which component, $a$ or $b$, those particles actually belong.

Correspondingly, equation~(\ref{eq:9}) for $z'$ can be rewritten in the following convenient form:
\begin{equation}
z' = {z_S}X , \label{eq:3_5}
\end{equation}
where $z_S$ is the convergence radius [see equation~(\ref{eq:6})] of the activity series in powers of $z'$ and the dimensionless variable, 
\begin{equation}
X = \frac{{{Z_a}}}{{{Z_b}}} , \label{eq:3_6}
\end{equation}
may take the values from $0$ to $1$ in order to ensure the series convergence ($z' < z_S$).

As a result, Mayer's expansions for a lattice model of binary mixtures takes the form,
\begin{equation}
\left. \begin{array}{l}
\displaystyle\ln {Z_b} = - \frac{1}{{{\rho _0}}}\sum\limits_{n \geqslant 1} {{b_n}{{\left( {{z_S}X} \right)}^n}} \\
\displaystyle {n_a} = \frac{1}{{{\rho _0}}}\sum\limits_{n \geqslant 1} {n{b_n}{{\left( {{z_S}X} \right)}^n}} 
\end{array} \right\} \left( 0 \leqslant X < 1 \right). \label{eq:3_7}
\end{equation}

In the divergence region of equation~(\ref{eq:3_7}) (i.e., at $ X > 1$), the corresponding ``high-density'' expansions of equation~(\ref{eq:2}) can be used, where $P$ and $\rho$ are again replaced by $ \ln {Z_b}$ and $n_a$, respectively, and the reciprocal activity of lattice-gas models [see equation~(\ref{eq:8})] is replaced by its analogy,
\[
\eta ' = {z_S}\frac{1}{X} ,
\]
for mixtures:
\begin{equation}
\left. \begin{array}{l}
\displaystyle\ln {Z_b} = - \ln X - \frac{1}{{{\rho _0}}}\sum\limits_{n \geqslant 1} {{b_n}{{\left( {\frac{{{z_S}}}{X}} \right)}^n}} \\
\displaystyle {n_a} = 1 - \frac{1}{{{\rho _0}}}\sum\limits_{n \geqslant 1} {n{b_n}{{\left( {\frac{{{z_S}}}{X}} \right)}^n}} 
\end{array} \right\} \left( X > 1 \right). \label{eq:3_8}
\end{equation}

The usage of the dimensionless parameters $n_a$ (or $n_b$), $Z_b$, $Z_a$, as well as the dimensionless variable $X$ in equations~(\ref{eq:3_7}), (\ref{eq:3_8}) significantly simplifies the transition from models of lattice gases to the corresponding models of mixtures and makes the theoretical curves universal for different models: all the specific parameters of the interaction potentials $ \phi _{aa}$, $ \phi _{ab}$, $ \phi _{bb}$ are actually incapsulated in the $Z_b$, $Z_a$ values and reduced potential, $ \phi ' $ (i.e., in the $\{ b_n \}$ set).

\begin{figure}
\centerline{\includegraphics[width=0.5\textwidth]{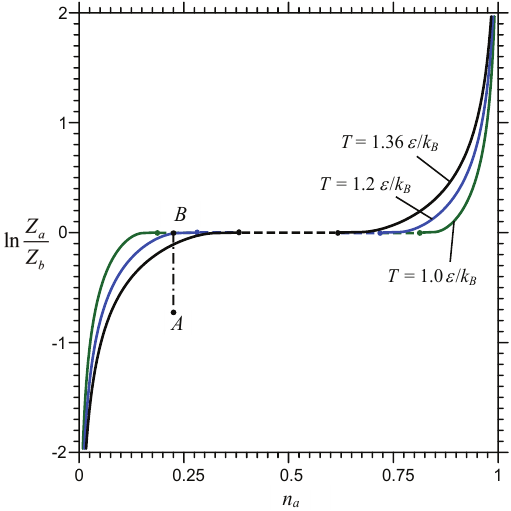}}
\caption{(Colour online) Dependences of the $ X = {Z_a}/{Z_b}$ parameter on the relative concentration of $a$ component, $n_a$, in the 3D cube-lattice model of binary mixtures at various subcritical temperatures.}
\label{fig:3_1}
\end{figure}

In particular, figure~\ref{fig:3_1} demonstrates the isothermal relation between ${n_a}$ and $ \ln {X}$ from equations~(\ref{eq:3_7}), (\ref{eq:3_8}) for the 3D cubic mixture model with the six n.n. attractions at several values of temperature. In the calculations, the $ \{ b^{\text{true}}_{n} \} $ set included $2000$ reducible cluster integrals evaluated on the basis of the first three irreducible integrals for this model (see table~\ref{tab:5}). The logarithmic dependence (i.e., using $\ln {X}$ instead of $X $) makes the isotherms absolutely symmetrical with respect to the $n_{a} = {1}/{2}$ point.

It is important to note that each subcritical isotherm of figure~\ref{fig:3_1} contains a horizontal interval (dashed lines in the figure) corresponding to the divergence regimes of equations~(\ref{eq:3_7}), (\ref{eq:3_8}) (when $X = 1$). This interval describes the two-phase states of the mixture at a given temperature, whereas the convergent regimes of equations~(\ref{eq:3_7}), (\ref{eq:3_8}) (left-hand and right-hand solid curves of each isotherm) correspond to the concentrations of single-phase states at the same temperature.

From this point of view, any isotherm describes the increase of the $n_a$ concentration (corresponding decrease of the $n_b$ concentration) in the mixture at constant temperature. In the low-concentration regimes, ${Z_a} < {Z_b}$ and the $a$ component remains completely dissolved in the $b$ component. When ${Z_a}$ reaches ${Z_b}$, the components tend to be separated from each other into two phases (the \emph{spinodal decomposition occurs}). Under the further increase of the $n_a$ concentration, one phase (where $b$ still prevails over $a$) decreases and the second phase (where $a$ became prevailing over $b$) increases in bulk while ${Z_a}$ stays equal to ${Z_b}$. At some concentration, $n_a$, high enough, ${Z_a}$ becomes prevailing over ${Z_b}$ --- the first phase with the prevailing $b$ component disappears and only the second phase with the prevailing $a$ component remains in the system --- the $b$ component becomes completely dissolved in the $a$ component. 

Similarly, one can consider the process of cooling (or heating) the mixture of some constant concentration, $n_a$ (see the dash-dotted line $AB$ in figure~\ref{fig:3_1}). At some high temperature (see the $A$ point), the mixture is in a single-phase state (where the $a$ component is dissolved in the $b$ component). Under the cooling, the temperature can reach some low value, where the corresponding isotherm has the horizontal interval at the given $n_a$ concentration (see the $B$ point in figure~\ref{fig:3_1}), and the spinodal decomposition occurs in the mixture. Under a further cooling, the system remains split into the two phases (the $B$ point does not change its position in the figure though it now belongs to other isotherms). Vice versa, the heating of the mixture can move the system from the $B$ point, i.e., lead to the dissolution of its components.

\begin{figure}[!t]
\centerline{\includegraphics[width=0.5\textwidth]{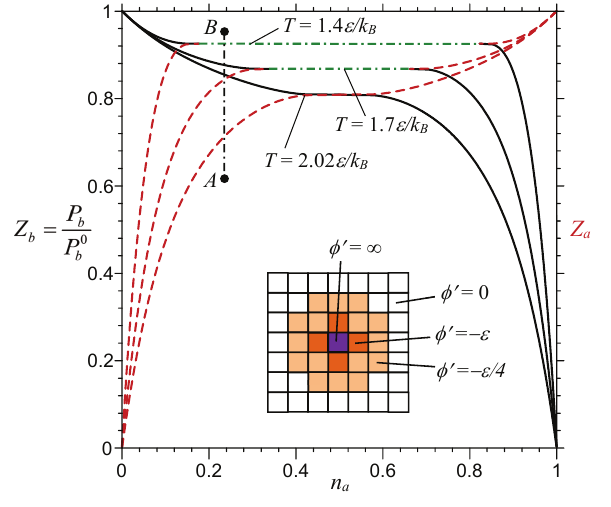}}
\caption{(Colour online) Isothermal $Z_b$ (solid lines) and $Z_a$ (dashed lines) dependences on the $a$ component molar fraction ($n_a$) for the two-dimensional ``long-range attraction'' mixture model. At each temperature, equations~(\ref{eq:3_7}), (\ref{eq:3_8}), (\ref{eq:3_9}) included the$ \{ b^{\text{true}}_{n} \} $ set of $10000$ reducible cluster integrals defined on the basis of the $ \{ \beta _{1}, \beta _{2}, \beta _{3}, \beta _{4} \} $ set for that model.}
\label{fig:3_2}
\end{figure}

As to the $ \ln {Z_b} $ expansions of equations~(\ref{eq:3_7}), (\ref{eq:3_8}), figure~\ref{fig:3_2} demonstrates three isothermal ${Z_b} ( {n_a} )$ dependences (solid curves) for the 2D square ``long-range attraction'' model of mixture (the isotherms of the corresponding lattice-gas model are shown in figure~\ref{fig:1_3}).

In case of perfect gaseous mixtures, a possible physical meaning for such ${Z_b}$ parameter was proposed by T.~Hill \cite{Hill}:
\[
{Z_{b}} = \frac{P_{b}} { P_{b}^0}, 
\]
where $ P_{b}$ is the actual partial pressure of gaseous component $b$ in the mixture and $ P_{b}^0$ is the pressure of pure $b$ vapor at the same conditions. Thus, the gaseous components may be completely mixed at some low or high concentrations (see curved intervals in figure~\ref{fig:3_2}) or split into two separate phases (see horizontal intervals in figure~\ref{fig:3_2}) depending on temperature. At some constant concentration, $n_a$, the components may be mixed at high temperature (see the $A$ point in figure~\ref{fig:3_2}) but separated at low temperature (see the $B$ point in the same figure).

As the $X$ variable defines the ${Z_a} / {Z_b}$ ratio [see equation~(\ref{eq:3_6})], equations~(\ref{eq:3_7}), (\ref{eq:3_8}) may also express the ${Z_a} ( X )$ dependence:
\begin{equation}
\left. \begin{array}{l}
\displaystyle\ln {Z_a} = \ln X - \frac{1}{{{\rho _0}}}\sum\limits_{n \geqslant 1} {{b_n}{{\left( {{z_S}X} \right)}^n}} {\rm{;}} \quad \quad 0 \leqslant X < 1 \\
\displaystyle\ln {Z_a} = - \frac{1}{{{\rho _0}}}\sum\limits_{n \geqslant 1} {{b_n}{{\left( {\frac{{{z_S}}}{X}} \right)}^n}} {\rm{;}} \quad \quad \quad \quad \quad \quad  X > 1 
\end{array} \right\} .\label{eq:3_9}
\end{equation}

As a result, the corresponding isothermal ${Z_a} ( {n_a} )$ dependences (dashed curves in figure~\ref{fig:3_2}) are always absolutely symmetrical to the ${Z_b} ( {n_a} )$ ones.

\subsection{Universality in the behavior of different lattice models}

As it is shown in previous sections, the known geometry and dimensionality of the lattice along with a certain function, $ \phi \left ( {\mathbf{r}_i}, {\mathbf{r}_j} \right ) $, must completely define the infinite set of reducible cluster integrals, $\left \{ b_{n} \left ( T \right ) \right \}$, and, thus, describe the behavior of three different classes of lattice models: i) lattice gases [see equations~(\ref{eq:1}), (\ref{eq:2})], where the $ \phi \left ( {\mathbf{r}_i}, {\mathbf{r}_j} \right ) $ has the meaning of the interaction energy per each pair of molecules, $i$ and $j$ (the pairwise interaction potential); ii) the Ising model of ferromagnets [see equations~(\ref{eq:3_1}), (\ref{eq:3_3})], where the interaction energy per each pair of spins, $ \Phi \left ( {\mathbf{r}_i}, {\mathbf{r}_j} \right ) = \phi \left ( {\mathbf{r}_i}, {\mathbf{r}_j} \right ) / 4$; iii) binary mixtures [see equations~(\ref{eq:3_7}), (\ref{eq:3_8})], where $ \phi \left ( {\mathbf{r}_i}, {\mathbf{r}_j} \right ) $ is equivalent to the reduced interaction potential, $ \phi ' \left ( {\mathbf{r}_i}, {\mathbf{r}_j} \right ) $ [see equation~(\ref{eq:8})].

In other words, if one studies the isotherms of some lattice gas, he or she can easily convert them to the magnetization curves of the equivalent magnetic model as well as the ${Z_b} ( {n_a} )$ [or ${Z_a} ( {n_a} )$] dependences for the corresponding mixture model. Therefore, any observed regularity in the behavior of distinct lattice gases would immediately mean the same regularity in the behavior of magnetics and mixtures, so let us hereafter consider the behavior of a certain class of lattice models (say, the magnetic models) and then expand the results to the other classes (i.e., lattice gases and mixtures). 

First of all, it should be emphasized that the presented approach to the evaluation of the unlimited $\left \{ b_{n} \left ( T \right ) \right \}$ sets (see equation~(\ref{eq:7}) and related paragraphs) is \emph{applicable at subcritical temperatures exclusively} (and, hence, for systems, where the critical point/Curie point exists). Only at subcritical values of temperature, the attractive interactions may prevail over the repulsive ones in $\exp \left ( \phi / [ {k_\text{B}} T] \right ) $ (in case of mixtures, the $a$-$a$ and $b$-$b$ attractions may prevail over the $a$-$b$ one in $\exp \left \{ \phi ' / [ {k_\text{B}} T] \right \}$), the reducible cluster integrals form a positive infinite set (in high orders at least), the corresponding power series diverge at some $z_S$ point, equation~(\ref{eq:5a}) has a positive root, $ {\rho _S} ( {z_S} ) < {\rho _0}/2$, and the asymptotic behavior of high-order reducible integrals can be expressed in terms of this $z_S$ [see equation~(\ref{eq:7})]. In case of magnetics, it means that the spontaneous magnetization is possible [$I \left ( H \to 0 \right ) = \pm {I_m} \neq 0$]. In case of lattice gases (or mixtures), there exists a jump of density (or concentration) at the $z_S$ point.

At supercritical temperatures, the $\left \{ b_{n} \left ( T \right ) \right \}$ sets are sign-changing and the activity power series are converging --- the present approach becomes inapplicable to the evaluation of such $\left \{ b_{n} \left ( T \right ) \right \}$ sets. Moreover, this approach can hardly be used even at the vicinity of the critical point itself. As the value of temperature comes close to the critical one, the behavior of cluster integrals becomes complicated which, in turn, fatally affects the actual accuracy of the $\left \{ b^{\text{true}}_{n} \right \}$ set approximated by using equation~(\ref{eq:7}) --- in order to achieve a higher accuracy there, the initial $\left \{ \beta _k \right \}$ sets should include much larger numbers of irreducible integrals than those presented in tables~\ref{tab:2}, \ref{tab:3}, \ref{tab:4}, \ref{tab:5}, \ref{tab:6}, \ref{tab:7}. 

For the Lee-Yang model (see figures~\ref{fig:1_2}, \ref{fig:2_1}), computations demonstrate that the varying of $k_\text{max}$ from $1$ to $6$ yields the deviations of results up to $4 \div 6\% $ at the values of temperature sufficiently lower than the critical one ($ T \leqslant 0.9 {T_C}$) though, at higher values of temperature (closer to ${T_C}$), the deviations become much larger (up to $10 \div 15\% $). 

Unfortunately, this situation makes it difficult to determine the critical temperature for other lattice models (for example, the ${T_C}$ of the ``long-range attraction'' model remains unknown at the moment). Based on equations~(\ref{eq:3_1}), (\ref{eq:3_3}) with the really accurate $\left \{ b_{n} \left ( T \right ) \right \}$ set, the spontaneous magnetization, $I_m$, must theoretically tend to $0$ exactly at $T \to {T_C}$. In case of lattice gases (or mixtures), the left-hand branch of the critical isotherm must meet the left-hand branch at $ {\rho } = {\rho _0}/2$ (or $ {n_a} = 1/2$, see the lowest isotherms in figure~\ref{fig:3_2}). 

As the actual $\left \{ b_{n} \left ( T \right ) \right \}$ sets are very approximate at the vicinity of $ T_C$, any practical evaluation of the critical point by using the above-mentioned criterion ($I_m \to 0$ at $T \to {T_C}$) yields the resulting temperature considerably higher than the true $ T_C$ (let us hereafter refer to such inaccurate critical temperature as ``pseudo-critical'', $T_0$). For example, the varying of $k_\text{max}$ from $1$ to $4$ for 3D cubic model with six nearest-neighbors (see table~\ref{tab:5}) changes such ``pseudo-critical'' temperature in the ranges from $1.53 \varepsilon / {k_\text{B}}$ to $1.51 \varepsilon / {k_\text{B}}$, whereas the numerically established critical temperature of this model ($T_C \approx 1.12787891 \varepsilon / {k_\text{B}}$ \cite{Ferrenberg, CMP2, CMP3}) is actually $25 \%$ lower.

\begin{table}[htb]
\noindent\caption{\label{tab:8} Pseudo-critical temperature, $T_0$, estimated for various models (for approximated virial series, the value of spontaneous magnetization reaches zero at this temperature).}
\begin{center}
\renewcommand{\arraystretch}{1.5}

\begin{tabular}{|| @{\hspace{20pt}}l @{\hspace{10pt}}| @{\hspace{20pt}} l @{\hspace{20pt}}||}

\hline 
\hline 
   Model& $T_0$ (in $\varepsilon / {k_\text{B}}$ units) \\[3pt]
\hline 
   2D Lee-Yang square model (4 neighbors) & $0.98$     \\[3pt]
   2D square model (8 neighbors) & $1.89$     \\[3pt]
   2D triangle model (6 neighbors) & $1.4868$     \\[3pt]
   3D cubic model (6 neighbors) & $1.51$     \\[3pt]
   3D cubic model (26 neighbors) & $6.7$     \\[3pt]
   2D ``long-range attraction'' model & $2.02$     \\[3pt]
\hline 
\hline 
\end{tabular}

\renewcommand{\arraystretch}{1.0}
\end{center}
\end{table}

Despite the obvious inaccuracy of such "pseudo-critical" temperature, $T_0$, it can easily be estimated in calculations for various models (see Table~\ref{tab:8}), thus providing a convenient way to define the \emph{reduced temperature}, ${T^*} = T / {T_0}$, in search for some universal parameters, because the dimensionless expressions for $I \left ( H \right ) $ in equations~(\ref{eq:3_1}) and (\ref{eq:3_3}) are expected to behave similarly for different models at the same values of ${T^*}$: $I \left ( H \to 0 \right ) \to 0$ at ${T^*} \to 1$; $I \left ( H \to 0 \right ) \to 1$ at ${T^*} \to 0$; $I \left ( H \to \infty \right ) \to 1$ at any ${T^*}$.

\begin{figure}[!t]
\centerline{ \includegraphics[width=0.5\textwidth]{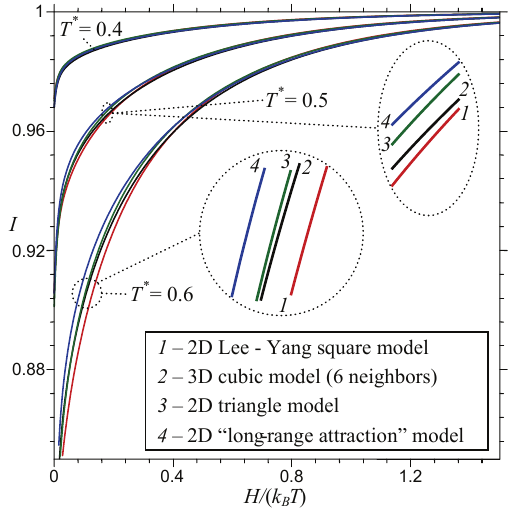} }
\caption{(Colour online) Magnetization curves of various models at different values of the reduced temperature, $ {T^*} = T/{T_0}$. The $ \{ b^{\text{true}}_{n} \} $ set of equation~(\ref{eq:3_3}) included $1000$ reducible cluster integrals calculated on the basis of the first three irreducible integrals for each model, $\left \{ {\beta _1}, {\beta _2}, {\beta _3} \right \}$.}
\label{fig:4_1}
\end{figure}

Indeed, the performed calculations have confirmed such qualitative and even quantitative similarity for the models of various geometry and dimensionality with different interaction potentials (see figure~\ref{fig:4_1}). In order to simplify that figure and to improve the visibility, the curves of only four models from table~\ref{tab:8} are shown in figure~\ref{fig:4_1}, but, in fact, the curves of the other models well agree with the presented ones. 

\begin{figure}[!t]
\centerline{\includegraphics[width=0.5\textwidth]{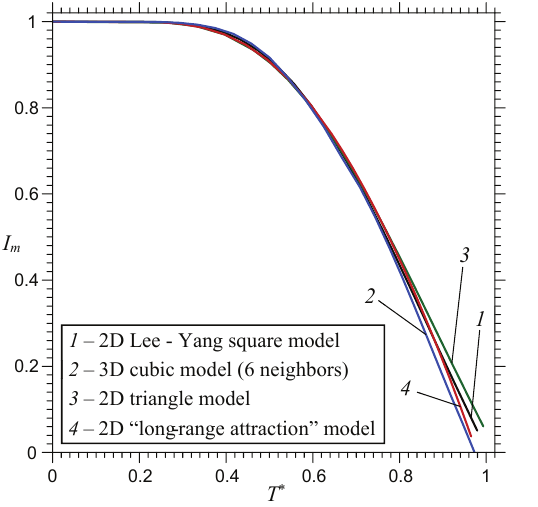} }
\caption{(Colour online) Spontaneous magnetization curves, ${I_m} ({T^*})$, of various lattice models. All the parameters of calculations are the same as for figure~\ref{fig:4_1}.}
\label{fig:4_2}
\end{figure}

In terms of the reduced temperature, $ {T^*}$, the loci of spontaneous magnetization points, $I_{m} (T)$, have turned out to be also very similar for fundamentally different models of ferromagnets (see figure~\ref{fig:4_2})

Some small deviations in the curves of figure~\ref{fig:4_2} at the vicinity of ${T^*} = 1$ are due to the difference in the number of irreducible integrals used in the calculations of the curves ($k_\text{max} = 3$ for all models) and estimations of ${T^*}$ itself (the maximum number of irreducible integrals known for each model).

As it is stated above, such quantitative universality in magnetization of fundamentally different lattice models of ferromagnets immediately means the universality in phase-transitions of different lattice-gas models and binary mixtures. This universality would not be surprising for mean-field approximations, where, in fact, neither dimensionality nor interaction potential matter at a certain choice of the mean-field model parameters. As to the considered cluster expansions, where the power coefficients (cluster integrals) fundamentally differ depending on the mathematical form of the interaction potential as well as the actual geometry and dimensionality of the model, the observed quantitative similarity deserves a further attentive study and, in future, may considerably simplify the corresponding theoretical predictions for some models remaining unexplored at the moment.

Though the presented here approach to approximate the subcritical $\left \{ b_{n} \left ( T \right ) \right \}$ sets remains inapplicable to the systems, where the first-order phase transitions cannot occur even at low temperatures (for example, the repulsion always prevails over attraction in \emph{antiferromagnets} or the $a$-$b$ attraction may always prevail over $a$-$a$ and $b$-$b$ ones in some mixtures/alloys), the obtained results can nevertheless be useful for describing the so-called \emph{order-disorder phase transitions} in such systems (for example, the well-known transitions in brass as an alloy of copper and zinc).

Indeed, the \emph{degree of long-range order}, $s$, of a certain antiferromagnet [with some hardcore interaction potential, ${\Phi _0} (r)$: ${\Phi _0} (0) = \infty$] is exactly the same as the spontaneous magnetization, $I_m$, of the corresponding ferromagnet [with the hardcore interaction potential, ${\Phi} (r)$, which is opposite to ${\Phi _0} (r)$ in all the coordination spheres beginning from the first one : ${\Phi} (0) = \infty$; ${\Phi} (r > 0) = - {\Phi _0} (r)$] at the same temperature \cite{Hill}. 

In other words, the temperature dependences of the long-range order degree, $s (T)$ for various antiferromagnets are actually described by the curves presented in figure~\ref{fig:4_2} for various ferromagnets: as the temperature increases, the degree of long-range order decreases reaching $0$ at $T_C$ (here, the $T_C$ value should be associated with the temperature, above which no long-range order can exist). Similarly, the same curves of figure~\ref{fig:4_2} can describe the order-disorder transitions in mixtures or alloys.

\section{Conclusions}

Modern achievements in Mayer's cluster expansion of the partition function have significantly advanced the statistical theory of phase-transitions --- a long-standing challenging problem of statistical mechanics and related fields (molecular physics, physical chemistry, etc.). In particular, a new approach to approximate the unlimited set of Mayer's cluster integrals (in practice, to the orders of thousands) make the theoretical description of phase transitions correct qualitatively as well as \emph{quantitatively}. 

In particular, we can conclude about the high efficiency of such approximation approach as applied to various statistical lattice models of matter, because the main parameter determining this approximation --- the phase-transition activity $z_S$ --- can be exactly expressed analytically for these models [see equation~(\ref{eq:6})]. 

Moreover, the recent reformulation of the mathematical relationship (see table~\ref{tab:1}) among the Ising problem, statistical lattice models of gases and binary mixtures (alloys, solutions) has finally allowed to eliminate the long-standing restriction traditionally inherent in such models --- the simplification of nearest-neighbor (n.n.) or next-nearest-neighbor (n.n.n.) interactions (when any system particle can interact to others only in the closest coordination spheres with some constant energy) --- thus, making the lattice models more realistic and closer to various real systems. Due to this reformulation we may now speak about a general ``lattice statistics'' instead of the old term of ``nearest-neighbor lattice statistics'' (see~\cite{Hill}).

As a result, the phase-transitions can now be described with high accuracy for fundamentally different systems: condensation phenomena in lattice gases (see figures~\ref{fig:1_2}, \ref{fig:1_3}); spontaneous magnetization of ferromagnets (see figures~\ref{fig:2_1},\ref{fig:4_1},\ref{fig:4_2}); spinodal decomposition phenomena in binary mixtures (see figures~\ref{fig:3_1}, \ref{fig:3_2}). It is important to note that the mentioned approximation approach is based on the information about only several low-order irreducible cluster integrals or virial coefficients (see tables~\ref{tab:2}, \ref{tab:3}, \ref{tab:4}, \ref{tab:5}, \ref{tab:6}, \ref{tab:7}) and it does not require any additional data (for example, empirical ones).

Another important conclusion concerns the observed universality in the behavior of different lattice systems: when expressed in terms of some dimensionless parameters, the isotherms of lattice gases, equilibrium curves of mixtures, and magnetization curves of ferromagnets are very similar qualitatively and even quantitatively for the lattices of various geometry and dimensionality with different interaction potentials at the same values of a certain reduced temperature (see figures~\ref{fig:4_1}, \ref{fig:4_2}). This universality deserves a further study and, potentially, it may considerably simplify theoretical predictions for some lattice models unexplored yet.

Though, strictly saying, the discussed approach to approximate the $\left \{ b_{n} \left ( T \right ) \right \}$ sets is applicable at subcritical values of temperature only (i.e., for lattice models, where the first-order phase transitions can occur), there is also a possibility to expand the presented results to the quantitative description of fundamentally different phase transitions in other systems (such as the order-disorder transitions in antiferromagnets and some specific mixtures/alloys).

\section*{Acknowledgements}

M. Ushcats and S. Ushcats thank US Government and Science \& Technology Center in Ukraine for funding (NSEP F005 1026).


\bibliographystyle{cmpj}
\bibliography{mainRev}

%
%
\newpage
\ukrainianpart

\title{Успіхи групового підходу Майєра в кількісному теоретичному описі фазових перетворень для різноманітних ґратчастих моделей матерії}
\author{М. В. Ушкац\refaddr{addr1}, Л. А. Булавін\refaddr{addr2}, С. Ю. Ушкац\refaddr{addr1}, Ж. Ю. Буруніна\refaddr{addr1}, О. В. Майборода\refaddr{addr1}, Н. О. Романчук\refaddr{addr1}, Н. О. Шаповал\refaddr{addr1}}
\addresses{
\addr{addr1} Національний університет кораблебудування ім. адмірала Макарова, просп. Героїв України, 9, м.~Миколаїв, 54025, Україна
\addr{addr2} Київський національний університет ім. Тараса Шевченка, просп. Академіка Глушкова, 2, м.~Київ, 03680,~Україна
} 
%
%
%

\makeukrtitle

\begin{abstract}
\tolerance=3000%
Останні досягнення в статистичній теорії, а саме можливість відтворювати майже необмежені ряди за степенями активності в розкладі Майєра на основі інформації про їх радіус збіжності, з одного боку, та узагальнення статистичних ґраткових моделей завдяки усуненню традиційного спрощення взаємодії лише найближчих сусідів, з іншого боку, дозволили точно кількісно описувати конденсацію в ґраткових газах, намагніченість феромагнетиків та спінодальний розпад в бінарних сумішах шляхом визначення лише декількох незвідних групових інтегралів (віріальних коефіцієнтів). Зокрема, результати розрахунків свідчать про якісну і навіть кількісну універсальність у поведінці згаданих ґраткових систем різної геометрії та вимірності за однакових значень приведеної температури, якщо цю поведінку виражати в термінах певних безрозмірних параметрів. У статті також обговорюється додаткова можливість опису фазових переходів порядок-безлад в інших ґраткових системах (наприклад, антиферомагнетиках і сплавах).
\keywords {ґраткова модель, розклад Майєра, груповий інтеграл, конденсація, намагнічування, спінодальний розпад}

\end{abstract}

\end{document}